\documentclass{PoS}
        \usepackage[tbtags]{amsmath}
	\usepackage{amssymb}

        \title{$b \to s\ell^+\ell^-$ in the high $q^2$ region at two-loops}
        
        \ShortTitle{$b \to s\ell^+\ell^-$ in the high $q^2$ region at two-loops}
        
        \author{\speaker{Volker Pilipp}, Christof Sch\"upbach%
                 \\
                Albert Einstein Center for Fundamental Physics\\
Institute for Theoretical Physics, University of Bern, Sidlerstrasse 5, 3012 Bern, Switzerland\\
                E-mail: \email{volker.pilipp@itp.unibe.ch}, \email{christof.schuepbach@itp.unibe.ch}}
        
        \abstract{We report on the first analytic NNLL calculation for the matrix elements of the operators $O_1$ and $O_2$ for the inlusive process $b\to X_s l^+l^-$ in the kinematical
region $q^2>4m_c^2$, where $q^2$ is the invariant mass squared of the lepton-pair.}
        
        \FullConference{RADCOR 2009 - 9th International Symposium on Radiative Corrections (Applications of Quantum Field Theory to Phenomenology) \\
                         October 25-30 2009\\
                         Ascona, Switzerland}

\begin{document}
        
        \section{Introduction}

In the Standard Model, the flavor-changing neutral current
process $b\to X_s l^+ l^-$ only occurs at the one-loop level
and is therefore sensitive to new physics. In the kinematical
region where  the lepton invariant mass squared $q^2$ is far
away from the $c\bar{c}$-resonances, the dilepton invariant
mass spectrum and the forward-backward asymmetry can be
precisely predicted using large $m_b$ expansion, where the leading
term is given by the partonic matrix element of the effective
Hamiltonian 
\begin{equation}
  \mathcal{H}_{eff} = -\frac{4 G_F}{\sqrt{2}} V_{ts}^* V_{tb} \sum\limits_{i=1}^{10} C_i(\mu) O_i (\mu).
\end{equation}
We neglect the CKM combination $V_{us}^* V_{ub}$ and the operator
basis is defined as in \cite{Bobeth:1999mk}. 
In \cite{ours} we published the first analytic NNLL calculation of the
high $q^2$ region of the matrix elements of the operators
\begin{equation} \label{oper}
  O_1  =  (\bar{s}_L \gamma_\mu T^a c_L) (\bar{c}_L \gamma^\mu T^a b_L),\qquad 
  O_2  =  (\bar{s}_L \gamma_\mu c_L) (\bar{c}_L \gamma^\mu b_L) \, ,
\end{equation}
which dominate the NNLL amplitude numerically. 
Earlier these results were only available analytically in the region of low $q^2$ 
\cite{Asatryan:2001zw,Asatryan:2002iy}. 
Using equations of motion the NNLL matrix elements of the effective
operators take the form
\begin{equation}
\langle s\ell^+\ell^-|O_i|b\rangle_\text{2-loops} = 
-\left(\frac{\alpha_s}{4\pi}\right)^2
\left[
F_i^{(7)}\langle O_7\rangle_\text{tree}+
F_i^{(9)}\langle O_9\rangle_\text{tree}
\right],
\label{formfact}
\end{equation} 
where $O_7=e/g_s^2 m_b (\bar{s}_{L} \sigma^{\mu\nu} b_{R})F_{\mu\nu}$
and $O_9 = e^2/g_s^2 (\bar{s}_L\gamma_{\mu} b_L)\sum_l(\bar{l}\gamma^{\mu}l)$.

\section{Calculations}
\begin{figure}[hb]
\centerline{\includegraphics[width=\textwidth]{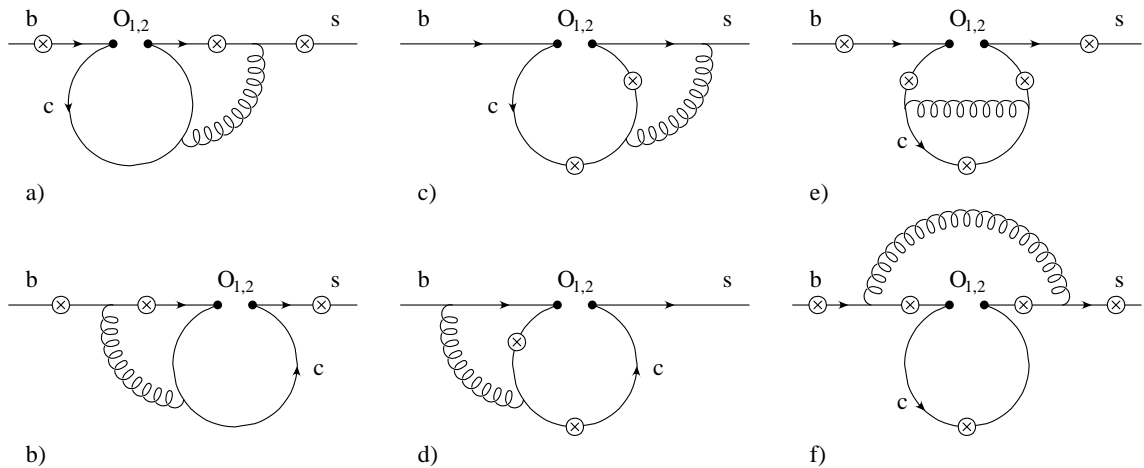}}
\caption{Diagrams that have to be taken into account 
  at order $\alpha_s$. The circle-crosses denote the possible
  locations where the virtual photon is emitted (see text).}
\label{f1}
\end{figure}
        
The diagrams contributing at order $\alpha_s$ are shown in Figure \ref{f1}. We set $m_s=0$ and define 
\begin{equation}
 \hat{s}=\frac{q^2}{m_b^2} \quad \mbox{and} \quad z=\frac{m_c^2}{m_b^2},
\end{equation}
where $q$ is the momentum of the virtual photon.
After reducing occurring tensor-like Feynman integrals \cite{Passarino:1978jh} the remaining scalar integrals can be further reduced to master integrals using integration by parts (IBP) identities \cite{CT}.
Considering the region $\hat{s}>4z$, we expanded the master
integrals in $z$ and kept the full analytic dependence in $\hat{s}$.

For power expanding Feynman integrals we use a combination of
\emph{method of regions} 
\cite{BGSS} 
and \emph{differential equation techniques} 
\cite{BKPPR,Pilipp:2008ef}:
Consider a set of Feynman integrals $I_1,\ldots,I_n$ depending on
the expansion parameter $z$ and related
by a system of differential equations obtained by differentiating $I_\alpha$ with respect to $z$ and applying IBP identities:
\begin{equation}
\frac{d}{dz}I_\alpha = \sum_\beta h_{\alpha\beta} I_\beta + g_\alpha,
\label{mi1}
\end{equation}
where $g_\alpha$ contains simpler integrals which pose no serious problems. 
Expanding both sides of (\ref{mi1}) in $\epsilon$, $z$ and $\ln z$
\begin{equation}
I_\alpha = \sum_{i,j,k} I_{\alpha,i}^{(j,k)} 
\epsilon^i z^j (\ln z)^k,\qquad
h_{\alpha\beta}=\sum_{i,j} h_{\alpha\beta,i}^{(j)}
\epsilon^iz^j,\qquad
g_\alpha=\sum_{i,j,k} g_{\alpha,i}^{(j,k)}
\epsilon^iz^j(\ln z)^k,
\label{mi2}
\end{equation}
and inserting (\ref{mi2}) into (\ref{mi1}) we obtain algebraic
equations for the coefficients $I_{\alpha,i}^{(j,k)}$
\begin{equation}
  0=
  (j+1)I_{\alpha,i}^{(j+1,k)}+(k+1)I_{\alpha,i}^{(j+1,k+1)}-
  \sum_{\beta}\sum_{i^\prime}\sum_{j^\prime}
    h_{\alpha\beta,i^\prime}^{(j^\prime)}
    I_{\beta,i-i^\prime}^{(j-j^\prime,k)}
    -g_{\alpha,i}^{(j,k)}.
  \label{mi3}
\end{equation}
This enables us to recursively calculate higher powers of $z$ once the leading powers are known. In practice this means that we need the
$I_{\alpha,i}^{(0,0)}$ and sometimes also the $I_{\alpha,i}^{(1,0)}$
as initial condition to (\ref{mi3}). These initial conditions can be computed using method of regions.
A non trivial check is provided by the fact that the leading terms containing logarithms of $z$ can be calculated by both
method of regions and the recurrence relation (\ref{mi3}).

The summation index $j$ in (\ref{mi2}) can take integer or
half-integer values, depending on the specific set of integrals $I_\alpha$.
In order to determine the possible powers of $z$ and $\ln(z)$ we used
the algorithm described in \cite{Pilipp:2008ef}. A given
$D$-dimensional $L$-loop Feynman
integral $I(z)$ reads in Feynman parameterization
\begin{equation}
I(z)=(-1)^N\left(\frac{i}{(4\pi)^{D/2}}\right)^L
\Gamma(N-LD/2)\int d^Nx\,\delta(1-\sum_{n=1}^N x_n)
\frac{U^{N-(L+1)D/2}}{(zF_1+F_2)^{N-LD/2}},
\label{1.2}
\end{equation}
where $U$, $F_1$ and $F_2$ are polynomials in $x_n$. 
Using Mellin-Barnes representation
(\ref{1.2}) can be cast into the following form
\begin{equation}
\begin{split}
I(z) = \,&(-1)^N\left(\frac{i}{(4\pi)^{D/2}}\right)^L
\frac{1}{2\pi i}\int_{-i\infty}^{i\infty}ds\,
z^s
\Gamma(-s)\Gamma(s+N-LD/2)\\
 \,&\times\int d^Nx\,\delta(1-\sum_{n=1}^N x_n)
U^{N-(L+1)D/2}F_1^sF_2^{-s-N+LD/2}.
\end{split}
\label{1.4}
\end{equation}
By closing the integration contour over $s$ to the right hand side the
poles on the positive real axis turn into powers of $z$. If we apply
the technique of \emph{sector decomposition} \cite{Binoth:2000ps} to 
(\ref{1.4}) we end up with terms of the following form
\begin{equation}
\sum_{l=1}^N\sum_k \int_0^1 d^{N-1}t\left(
\prod_{j=1}^{N-1}t_j^{A_j-B_j\epsilon-C_js}\right)
U_{lk}^{N-(L+1)D/2}F_{1,lk}^sF_{2,lk}^{-s-N+LD/2},
\label{1.6}
\end{equation}
where $U_{lk}$, $F_{1,lk}$ and $F_{2,lk}$ contain terms that are
constant in $\vec{t}$.
From (\ref{1.6}) we can read off that the poles in $s$ are located at:
\begin{equation}
s_{jn}=\frac{1+n+A_j-B_j\epsilon}{C_j},
\label{1.7}
\end{equation}
where $n\in \mathbb{N}_0$.
 
Additionally, the procedure described above
allows us to evaluate the coefficients of the expansion in $z$
numerically which we used to again test the initial conditions of the
differential equations.

\section{Results}
\begin{figure}
\begin{tabular}{l@{\hspace{-0.1cm}}l}
\resizebox{0.52\textwidth}{!}
{
\input{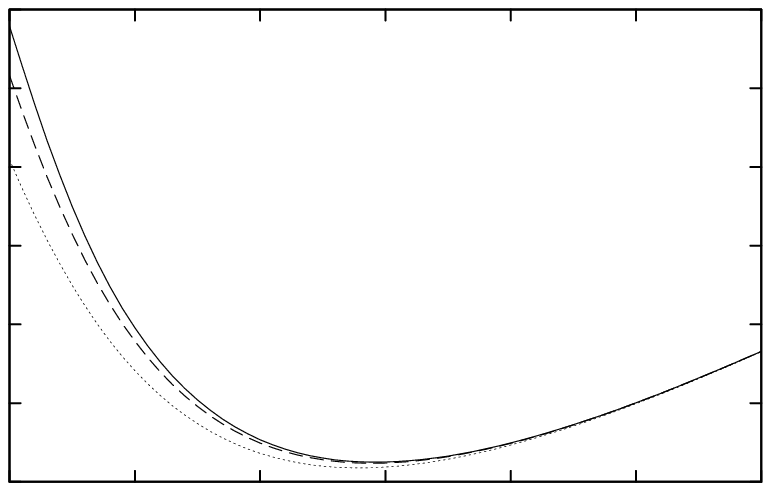}
}
&
\resizebox{0.52\textwidth}{!}
{
\input{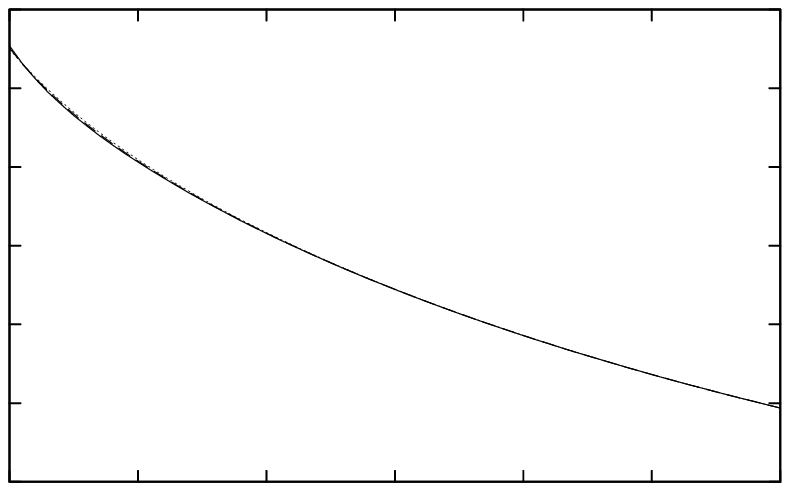}
}
\\[-0.1cm]
\resizebox{0.52\textwidth}{!}
{
\input{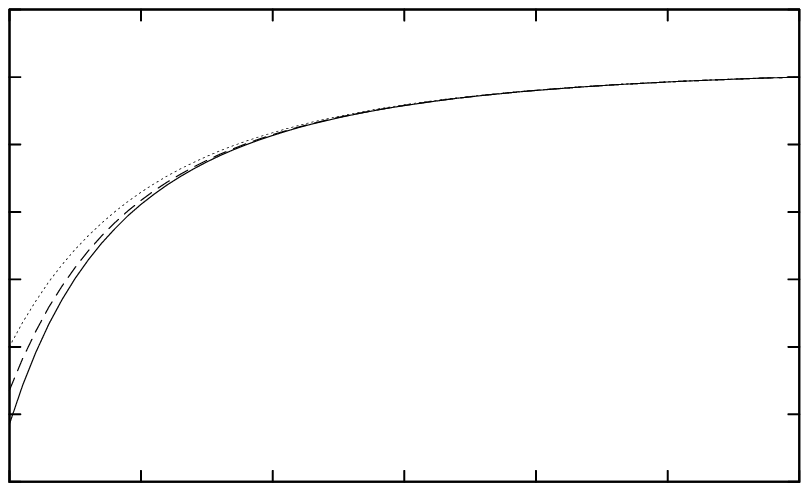}
}
&
\resizebox{0.52\textwidth}{!}
{
\input{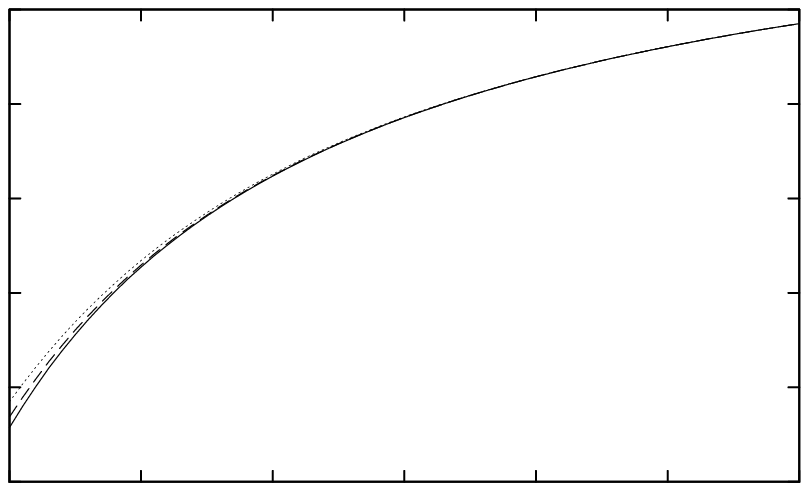}
}
\\[-0.1cm]
\resizebox{0.52\textwidth}{!}
{
\input{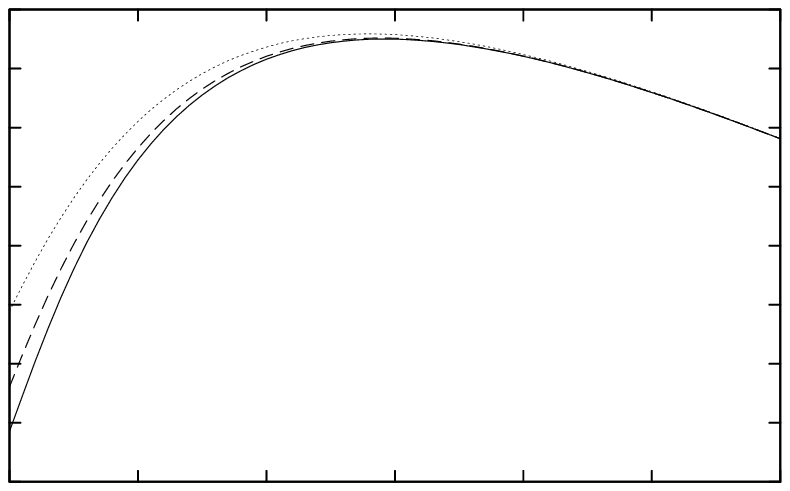}
}
&
\resizebox{0.52\textwidth}{!}
{
\input{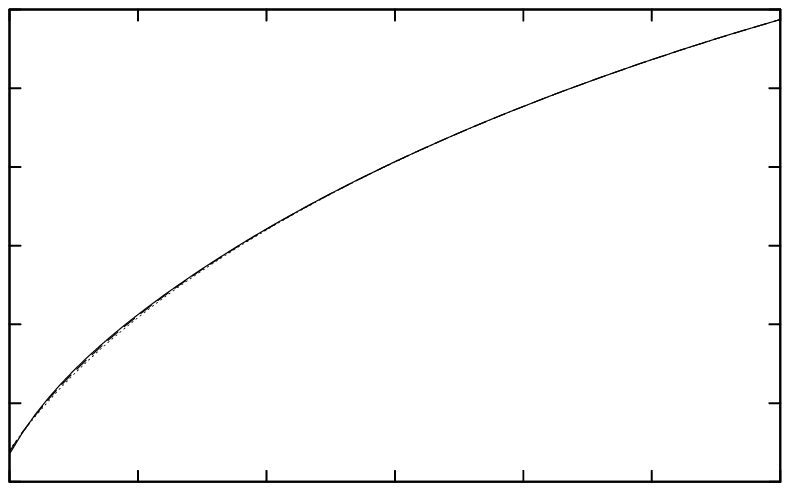}
}
\\[-0.1cm]
\resizebox{0.52\textwidth}{!}
{
\input{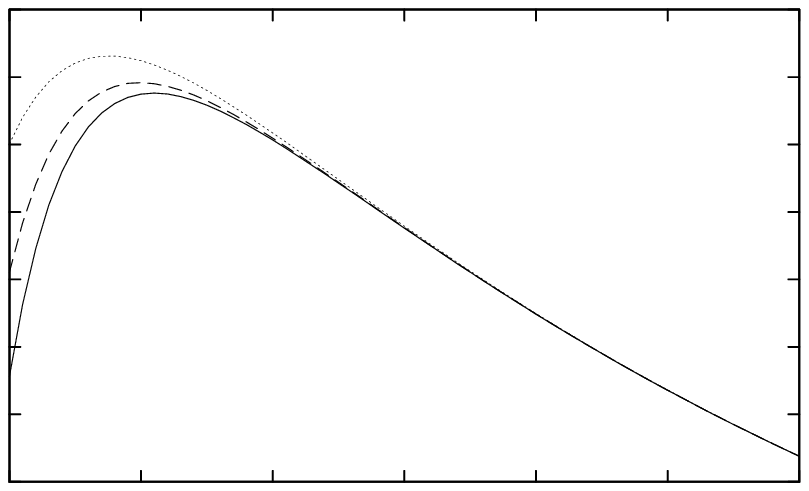}
}
&
\resizebox{0.52\textwidth}{!}
{
\input{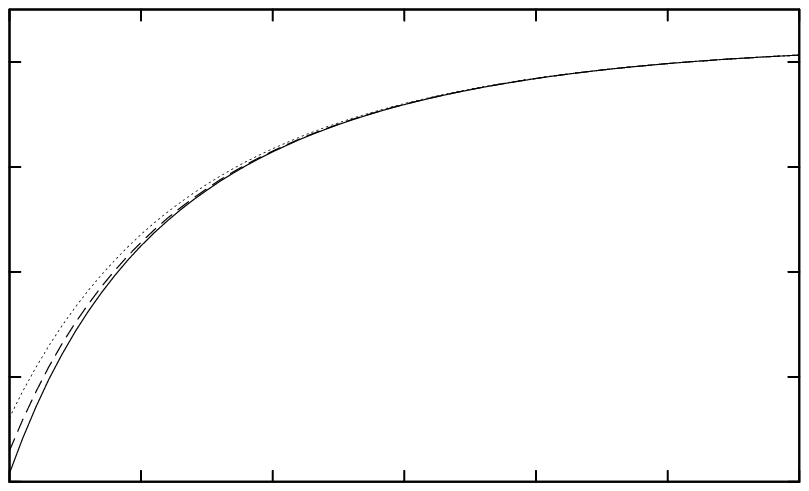}
}
\end{tabular}
\caption{Real and imaginary parts of the form factors
  $F_{1,2}^{(7,9)}$ as functions of
  $\hat{s}$. To demonstrate the convergence of the expansion in $z$ we
  included all orders up to $z^6$, $z^8$ and
  $z^{10}$ in the dotted, dashed and solid lines respectively. 
  We put $\mu=m_b$ and used the default value $z=0.1$.}

\label{ff}
\end{figure}
In order to get accurate results we keep terms up to $z^{10}$. Our
results agree with the previous numerical calculation 
\cite{Ghinculov:2003qd} within less than $1\%$ difference. 
To demonstrate the convergence of the power expansions,
we show in Figure~\ref{ff} the form factors defined in
(\ref{formfact}) as functions of $\hat{s}$,
where we include all orders up to $z^6$, $z^8$ and $z^{10}$. We use as
default value $z = 0.1$ such that the $c\bar{c}$-threshold is located at
$\hat{s} = 0.4$. One sees from the figures that far away from the 
$c\bar{c}$-threshold, i.e.\ for
$\hat{s} > 0.6$, the expansions for all form factors are well behaved.

\begin{figure}
\begin{minipage}[b]{0.5\linewidth}
\resizebox{\textwidth}{!}{\input{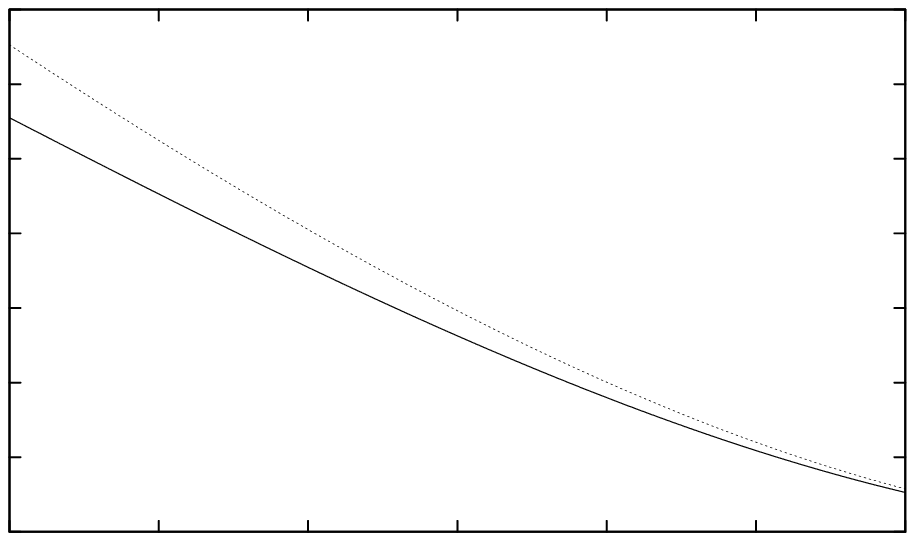}}
\end{minipage}
\begin{minipage}[b]{0.5\linewidth}
\resizebox{\textwidth}{!}{\input{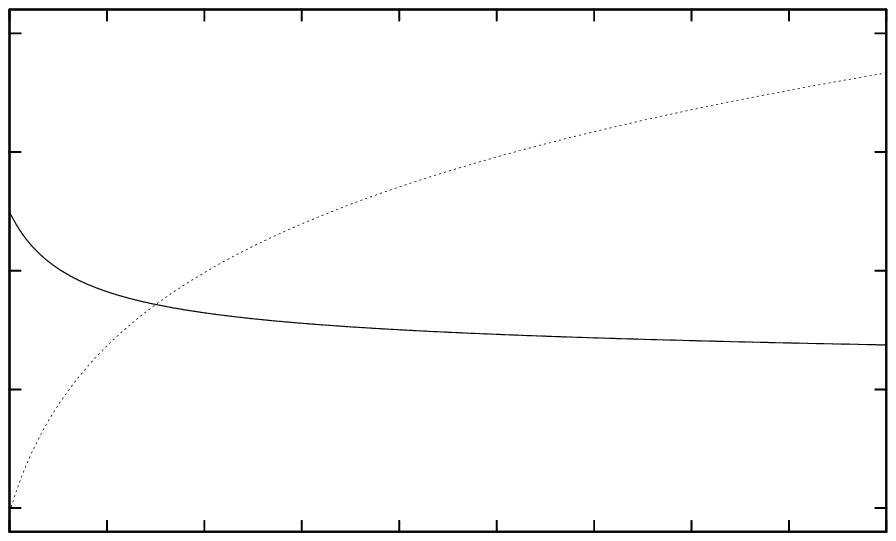}}
\end{minipage}
\caption{Perturbative part of
$R(\hat{s})$ (left) and $R_\text{high}$ (right) at NNLL. The solid lines represents the NNLL
result, whereas in the dotted lines the order $\alpha_s$ corrections 
to the matrix elements associated with $O_{1,2}$ are switched off.
In the left figure we use $\mu=m_b$. See text for details.}\label{f2}
\end{figure}

The impact of our 
results on the perturbative part of the high $q^2$-spectrum 
\cite{Asatryan:2001zw}
\begin{equation}
R(\hat{s})=\frac{1}{\Gamma(\bar{B}\to X_c e^- \bar{\nu}_e)}
\frac{d\Gamma(\bar{B}\to X_s\ell^+\ell^-)}{d\hat{s}}
\label{r1}
\end{equation}
is shown in Figure \ref{f2} (left), where we used the same parameters as in \cite{ours}.
The finite bremsstrahlung corrections calculated in 
\cite{Asatryan:2002iy} are neglected.
From Figure \ref{f2} (left) we conclude that for $\mu=m_b$ the contributions of our results 
lead to corrections of the order $10\% - 15\%$.
Integrating $R(\hat{s})$ over the high $\hat{s}$ region, we define
\begin{equation}
R_\text{high}=\int_{0.6}^1 d\hat{s}\, R(\hat{s}).
\end{equation}
Figure \ref{f2} (right) shows the dependence of the perturbative part of
$R_\text{high}$ on the
renormalization scale. 
We obtain 
\begin{equation}
R_\text{high,pert}=(0.43\pm0.01(\mu))\times 10^{-5},
\end{equation}
where we determined the error by varying $\mu$ between 2 GeV and 
10 GeV. The
corrections due to our results lead to a decrease of the scale dependence to $2\%$.

\acknowledgments
This work is partially
supported by the Swiss National Foundation, by EC-Contract
MRTNCT-2006-035482 (FLAVIAnet) and by the Helmholtz Association through 
funds provided to the virtual institute ''Spin and strong QCD'' 
(VH-VI-231).
The Albert Einstein Center for Fundamental Physics is supported by the 
''Innovations- und Kooperationsprojekt C-13 of the Schweizerische 
Universit\"atskonferenz SUK/CRUS''.

        \end{document}